\documentclass[aps,prl,twocolumn,showpacs,floatfix,superscriptaddress]{revtex4}
\usepackage{amsmath,amssymb,amsfonts,bbm,graphicx,hyperref,times}
\usepackage[dvips,usenames]{color}
\newcommand{\ket}[1]{\vert #1 \rangle}
\newcommand{\bra}[1]{\langle #1 \vert}
\def\Tr{\hbox{Tr}}
\begin{document}
\title{Optical phase estimation in the presence of  phase-diffusion}
\author{Marco G. Genoni}
\affiliation{QOLS, Blackett Laboratory, Imperial College London, London
SW7 2BW, UK}
\affiliation{Dipartimento di Fisica, Universit\`a degli Studi di Milano,
I-20133 Milano, Italy}
\author{Stefano Olivares}
\affiliation{Dipartimento di Fisica, Universit\`a degli Studi di
Trieste, I-34151 Trieste, Italy} 
\author{Matteo G. A. Paris}
\email{matteo.paris@fisica.unimi.it}
\affiliation{Dipartimento di Fisica, Universit\`a degli Studi di Milano,
I-20133 Milano, Italy}
\date{\today}
\begin{abstract}
The measurement problem for the optical phase has been traditionally 
attacked for noiseless schemes or in the presence of amplitude or detection 
noise. Here we address estimation of phase in the presence of phase diffusion 
and evaluate the ultimate quantum limits to precision for phase-shifted Gaussian 
states. We look for the optimal detection scheme and derive approximate scaling 
laws for the quantum Fisher information and the optimal squeezing fraction in terms 
of the total energy and the amount of noise. We also found that homodyne
detection is a nearly optimal detection scheme in the limit of very
small and large noise. 
\end{abstract}
\pacs{42.50.St,03.65.St,07.60.Ly}
\maketitle
The estimation of the optical phase in quantum mechanics is a longstanding 
problem with both fundamental and technological implications. The attempts to 
define a Hermitian  phase-operator conjugate to the number operator are the 
subject of an extensive literature \cite{rs} and 
several experimental protocols to estimate the value of the optical phase have 
been also proposed and demonstrated, in particular using different quantum
strategies and interferometric setups \cite{Armen02,Mitch04,Nag07,Res07,
Hig07,Hig09}, which have been shown to beat the standard quantum limit and to 
attain the Heisenberg limit \cite{Z92,Sam92,Giov046,Guo08}.
More recently the ultimate bounds to precision of phase estimation
with Gaussian states have been discussed in details
\cite{Mon06,Oli09} using local quantum estimation theory. 
Squeezed vacuum state has been shown to be the most sensitive
for a given average photon number and two adaptive local 
measurement schemes have been proposed to attain the Heisenberg limit
asymptotically.
\par
The estimation of the optical phase, besides the fundamental interest,  
is also relevant for optical communication scheme where information is 
encoded in the phase of laser pulses that must travel long distances 
between the sender and receiver. In such a context the receiver has to 
decode the information carried by the optical wave-packets which will be 
unavoidably degraded by different sources of noise, which have to be duly 
taken into account in the quantum estimation problem. As a matter of fact, 
only amplitude and/or detection noise have been taken into account in the 
analysis of quantum phase estimation, e.g. imperfect photodetection in the 
measurement stage, or amplitude noise in interferometric setups 
\cite{Par95,BanPhNat,BanPhPRL,Cab10,DurkinPh}.
The role of phase-diffusive noise in phase-estimation have been
investigated for qubit \cite{Bri10,Ber10} and in part for condensate 
systems \cite{BEC1,BEC2}, while no similar analysis have been done 
for a continuous variable system. Phase-diffusive noise is the most 
detrimental for phase-estimation since it destroys the off-diagonal
elements of the density matrix. Moreover, any quantum state that is 
unaffected by phase-diffusion, is also invariant under a phase-shift, 
and thus is totally useless for phase-estimation. 
\par
In this letter we address for the first time phase estimation in the presence of
phase diffusion, seek for the optimal Gaussian states, and evaluate
the ultimate quantum limits to precision of phase estimation. 
We also investigate whether the ultimate performances may be achieved
with feasible detection scheme and found that homodyne detection is nearly 
optimal for very small and large amount of noise.
\par
Let us start by a pico review of local quantum estimation theory
\cite{Mal9X,BC9X,Bro9X,LQE}. When a physical parameter is not directly accessible 
one has to resort to indirect measurements. Let us denote by $\phi$ the 
quantity of interest, $X$ the measured observable, and $\chi=(x_1, \ldots, 
x_M)$ the observed sample. The {\em estimation problem} amounts to find an 
estimator, that is a map $\hat{\phi}= \hat{\phi}(\chi)$ from the 
set of the outcomes to the space of parameters. Classically, optimal 
estimators are those saturating the Cram\'er-Rao inequality $
\textrm{Var}(\phi)\geq [M F(\phi)]^{-1}$
which bounds from 
below the variance $\textrm{Var}(\phi)=E
[\hat{\phi}^2]-E [\hat{\phi}]^2 $ of any unbiased estimator
of the parameter $\phi$. In the Cram\'er-Rao inequality, $M$ is the
number of measurements and $F(\phi)$ is the Fisher
Information (FI) $
F(\phi)= \int\! dx\, p(x|\phi)\left[
\partial_\phi \ln p(x|\phi) \right ]^2$
where $p(x|\phi)$ is
the conditional probability of obtaining the value $x$ when the
parameter has the value $\phi$.  The quantum analogue of the
Cram\'er-Rao bound is obtained starting from the Born rule $p(x|\phi)=
\Tr[\Pi_x \varrho_\phi]$ where $\{\Pi_x\}$ is the probability
operator-valued measure (POVM) describing the measurement and
$\varrho_\phi$ the density operator, labeled by the parameter of
interest. Upon introducing the Symmetric Logarithmic Derivative (SLD) 
$L_\phi$ as the operator satisfying $2 \partial_\phi\varrho_\phi=  
L_\phi \varrho_\phi+ \varrho_\phi L_\phi$ one proves that 
the FI is upper bounded by the Quantum 
Fisher Information (QFI) \cite{BC9X}
$F(\phi)\leq H(\phi)\equiv\Tr[\varrho_\phi L_\phi^2]$.
In turn, the ultimate limit to precision is given by the quantum 
Cram\'er-Rao bound $
\textrm{Var}(\phi)\geq [M H(\phi)]^{-1}$.
The family of states we are going to deal  
is a unitary one $\varrho_\phi= U_\phi \varrho_0 U_\phi^\dag=
\sum_k \lambda_k |\lambda_k(\phi)\rangle\langle\lambda_k(\phi)|$,
where $|\lambda_k(\phi)\rangle =U_\phi |\lambda_k\rangle$ and
$U_\phi=\exp\{-i \phi G\}$ describes a phase-shift with the 
single-mode number operator $G=a^\dag a$ as the generator.
In this case the SLD may be written 
as $L_\phi = U_\phi L_0 U_\phi^\dag$,
where $L_0$ is independent on $\phi$. 
The corresponding QFI does not depend on the parameter $\phi$, and 
reads
\begin{align}
H = \Tr [ \varrho_0 L_0^2 ]= 2 \sum_{n \neq m} 
\frac{(\lambda_n - \lambda_m)^2}{\lambda_n + \lambda_m} 
|\langle \lambda_n |G | \lambda_m\rangle | ^2 \label{eq:QFI}
\end{align}
\par
Phase-diffusion for a continuous-variable system is described by 
the master equation $\dot{\varrho} = \Gamma \mathcal{L}[a^{\dag} a] 
\varrho$ where $\mathcal{L}[O]\varrho = 2 O \varrho O^\dag - 
O^\dag O \varrho - \varrho O^\dag O$. The  solution for an initial 
state $\varrho(0)$ is given by
$ \varrho(t) = \mathcal{N}_{\Delta} (\varrho(0))  
= \sum_{n,m} e^{- \Delta^2 (n-m)^2} \varrho_{n,m}(0) |n\rangle\langle m|$
where $\Delta\equiv \Gamma t$, $\Gamma$ is the noise
amplitude and $\varrho_{n,m}(0)= \langle n | \varrho(0)|m\rangle$.
The diagonal elements $\varrho$ are left unchanged and, in turn,
energy is conserved, whereas the off-diagonal ones are progressively 
destroyed. 
\par
We assume that phase noise occurs between the application of the phase-shift 
and the detection of the signal, and address quantum estimation of a phase-shift 
applied to pure single-mode Gaussian states $|\psi_{\hbox{\scriptsize G}}\rangle = 
D(\alpha) S(r)  |0\rangle$ where $S(r)=\exp\{ (r/2)(a^2 - a^{\dag 2})$ is 
the squeezing operators, $D(\alpha) =  \exp\{ \alpha (a^\dag -  a)\}$ the 
displacement operator, being $r, \alpha \in \mathbbm{R}$.
The input state is firstly phase-shifted by applying the unitary operator
$U_\phi$, where $\phi$ is the unknown phase-shift,  
and then, before being measured, it undergoes phase-diffusion. 
Our aim is to determine the ultimate bound to precision for a generic 
pure Gaussian probe and then look for the optimal one by maximizing the 
QFI over the state parameters.
\par
The mixed non-Gaussian state that is being measured is  given by 
\begin{align}
\varrho_\phi (t) = \mathcal{N}_{\Delta} ( U_\phi |\psi_G\rangle\langle\psi_G| 
U_\phi^{\dag} )= U_\phi \mathcal{N}_{\Delta} ( |\psi_G\rangle\langle \psi_G|) 
U_\phi^{\dag}\notag\,,
\end{align}
where the second equality holds since the superoperator 
$\mathcal{L}[a^{\dag}a]$ and the phase-shift operator $U_\phi$ commute.
Because of this fact our estimation problem corresponds 
to the case of a unitary family described above, with the input 
mixed state given by $\mathcal{N}_{\Delta} ( |\psi_G\rangle\langle \psi_G|)$.
In order to evaluate the corresponding QFI one writes $\varrho_\phi$ in its 
diagonal form 
$\varrho_\phi = \sum_n \lambda_n \ket{\lambda_n(\phi)}\bra{\lambda_n(\phi)} =
\sum_n \lambda_n U_\phi\ket{\lambda_n}\bra{\lambda_n}U_\phi^{\dag}$,
where $|\lambda_n(\phi)\rangle$ and $|\lambda_n\rangle$ are respectively 
the eigenvectors of $\varrho_\phi$ and of 
$\mathcal{N}_{\Delta}(|\psi_G\rangle\langle\psi_G|)$
corresponding to the eigenvalues $\lambda_n$, which are in fact 
left unchanged by the phase-shift operation. 
By decomposing $|\lambda_n\rangle = \sum_k r_{nk} |k\rangle$
in the Fock basis and by substituting this into the eigenvalues equation 
$\mathcal{N}_\Delta(|\psi_G\rangle\langle\psi_G|) |\lambda_n\rangle 
= \lambda_n |\lambda_n\rangle$
we have $
\langle n | \psi_G\rangle\langle\psi_G | l\rangle e^{-\Delta^2 (n-k)^2} r_{qk} =
\lambda_q r_{qn}\, \forall\,n$. 
Moreover, since  $a^\dag a |\lambda_n \rangle  =
\sum_k k\, r_{nk}\, e^{i k \phi} \ket{k}$, we have that:
$|\langle \lambda_m \vert a^\dag a  \vert \lambda_n \rangle|^2 =
\left|\sum_k k\, r_{mk}\, r_{nk}\right|^2$.
After evaluating the QFI using the above formulas one sees that it
depends only on  the eigenvalues $\lambda_n$ and on the components of
the eigenvectors $r_{nk}$ which, being $\phi$ a unitary parameter, do
not depend on the parameter itself. The explicit 
values of $\lambda_n$ and $r_{nk}$ have been obtained by performing 
numerical diagonalization.
\par
Upon inspecting the solution of the master equation one sees that 
the vanishing of the off-diagonal matrix elements is governed by the
product between $\Delta^2$ and the squared  difference between the Fock
indices $(n-m)^2$.  Besides, for a pure Gaussian state, the presence of
non-zero (non-negligible) off-diagonal elements is somehow ruled by the
average photon-number $N=\langle a^{\dag} a \rangle$ and thus we roughly
expect the QFI to somehow depend on the quantity $\xi=N\Delta$. 
Pure Gaussian states may conveniently 
parametrized by the average photon number of photons $N$ and of the 
corresponding squeezing fraction $\beta$, in formula 
$N = \sinh^2 r + |\alpha|^2$ and $\beta = \sinh^2 r/N$, and thus 
the QFI will be function of the three parameters $N,\beta$ and $\Delta$.
\par
We start our analysis by evaluating the QFI at fixed noise $\Delta$. 
We consider four values of the maximum energy 
$N_{\max}=\langle a^{\dag}a \rangle_{\textrm{max}}= \{10,15,20,30\}$ 
(with $10$ steps on intermediate energies $N$) 
and different values of the noise parameter $\Delta$. The values 
of $\Delta$ are chosen such that we can find points corresponding to 
fixed values of $\xi$. The curves are built by looking for
the optimal pure Gaussian state, i.e. maximizing the QFI as a function
of the squeezing fraction $\beta$, for any fixed value of the energy 
$N$ and of the noise parameter $\Delta$.
\begin{figure}[h!]
\includegraphics[width=0.45\columnwidth]{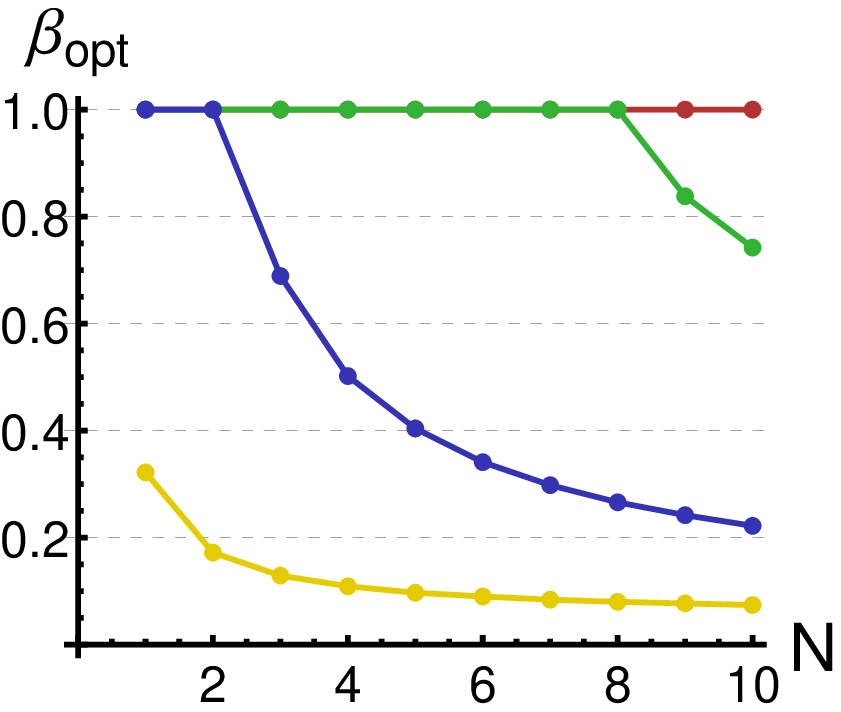}
\includegraphics[width=0.45\columnwidth]{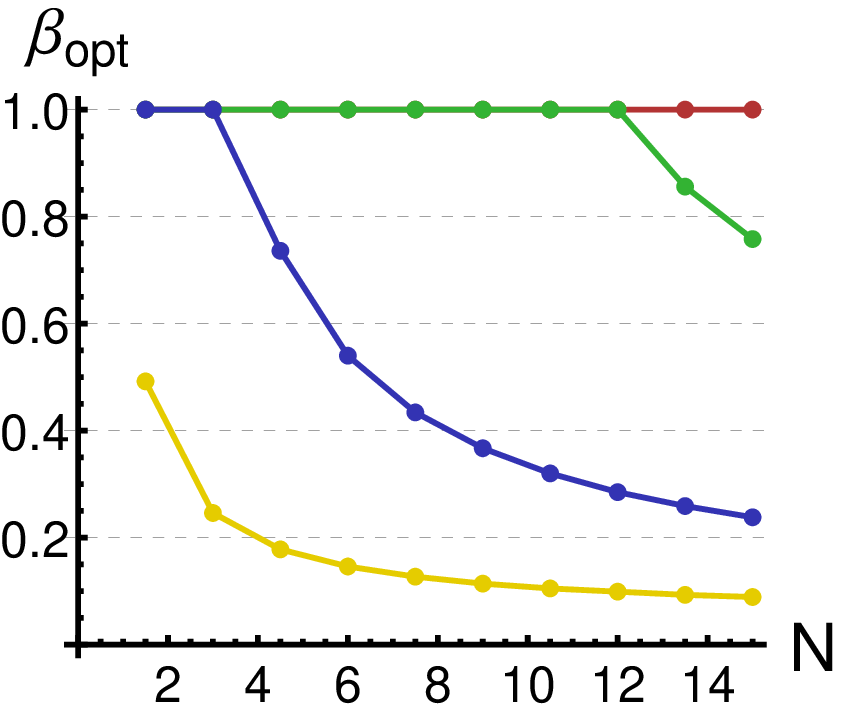}
\includegraphics[width=0.45\columnwidth]{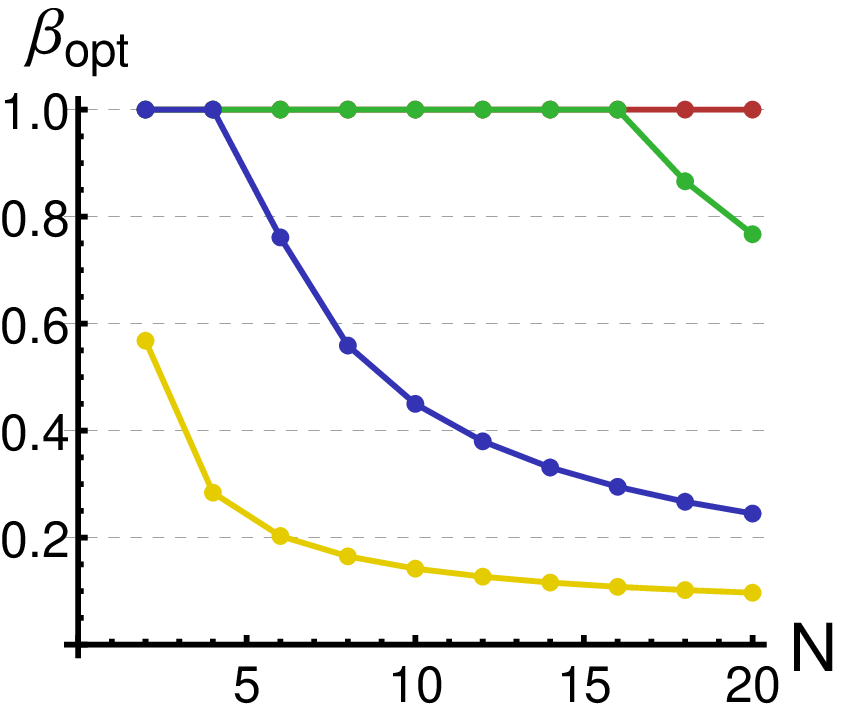}
\includegraphics[width=0.45\columnwidth]{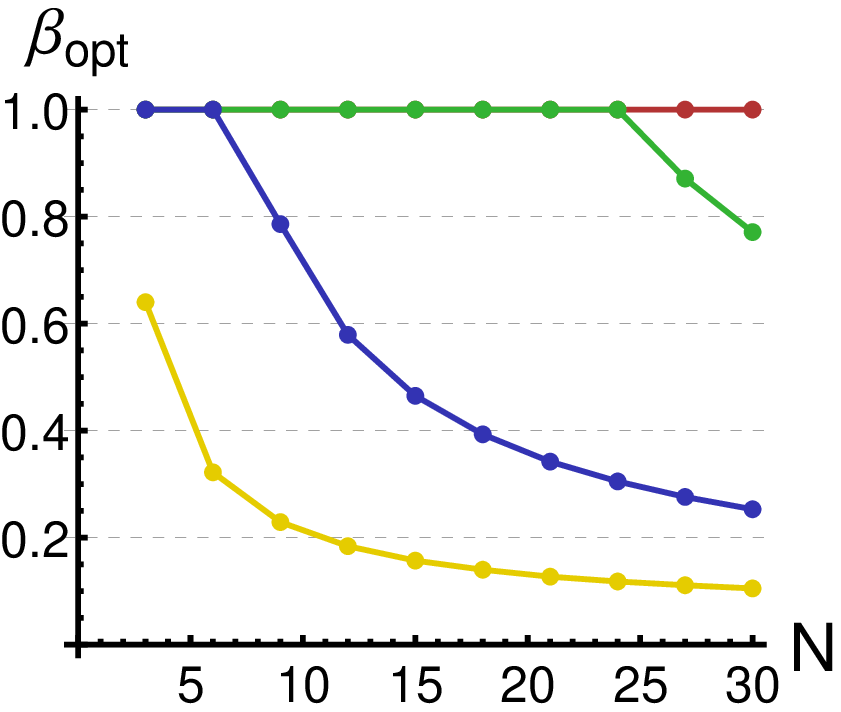}
\caption{
\label{f:optsqueezing} (Color online)
Optimal squeezing fraction $\beta$ as a function of 
the average photon number $N$ and for different values of $\Delta^2$.
(Top left): from top to bottom $\Delta^2=\{4.5\times 10^{-5}$, $4.5\times 10^{-4}$,
$4.5\times 10^{-3}$, $4.5\times 10^{-2} \}$. (Top right): from top to 
bottom $\Delta^2=\{2.0\times 10^{-5}$, $2.0\times 10^{-4}$,
$2.0\times 10^{-3}$, $2.0\times 10^{-2} \}$. (Bottom left): from top
to bottom $\Delta^2=\{1.125\times 10^{-5}$, 
$1.125\times 10^{-4}$, $1.125\times 10^{-3}$, 
$1.125\times 10^{-2} \}$. (Bottom right): from top to bottom
$\Delta^2=\{5.0\times 10^{-6}$, $5.0\times 10^{-5}$,
$5.0\times 10^{-4}$, $5.0\times 10^{-3} \}$.
}
\end{figure}
\par
\begin{figure}[h!]
\includegraphics[width=0.46\columnwidth]{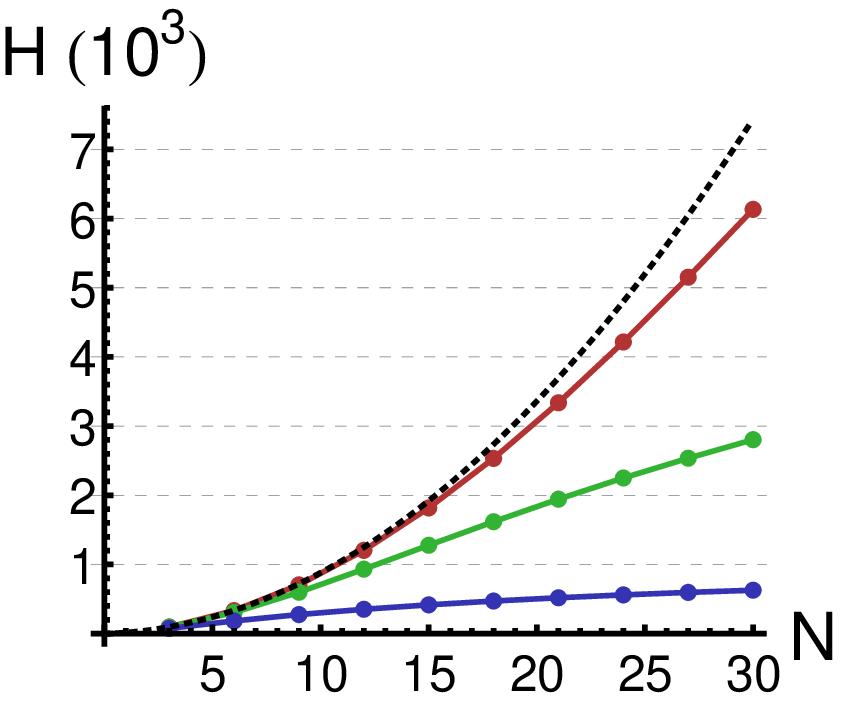}
$\:$\includegraphics[width=0.45\columnwidth]{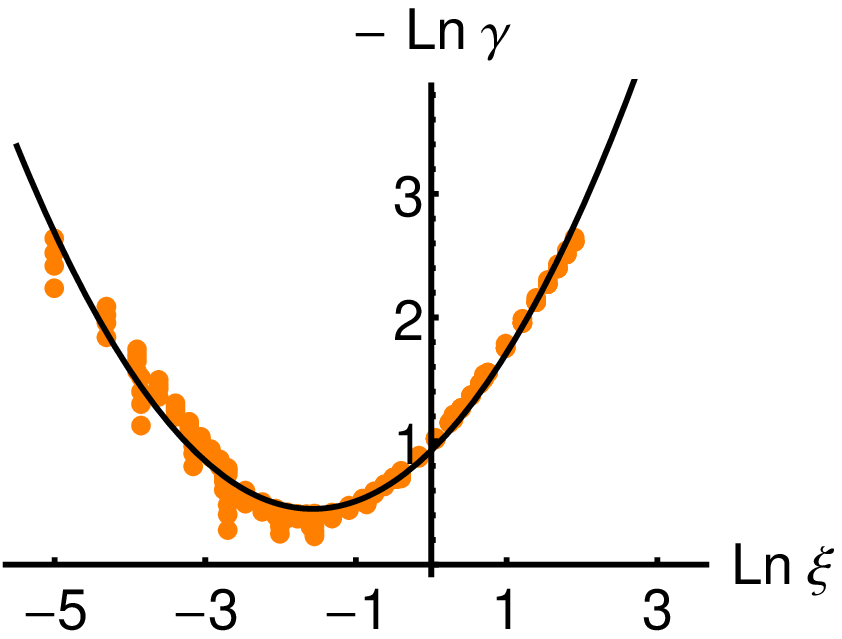}
\caption{
\label{f:QFIPhDiffCV} (Color online) Left panel: QFI 
of optimized pure input Gaussian states 
as a function of the average photon number $N$ 
and for different values of the noise parameter $\Delta$; 
from top to bottom $\Delta^2=\{5.0\times 10^{-6}$, 
$5.0\times 10^{-5}$, $5.0\times 10^{-4} \}$. The black 
dotted line is the QFI for the noiseless case $H(N,\beta=1,
\Delta=0) = 8(N^2 + N)$. 
Right panel: (orange points) $-\ln\gamma(\xi)$ as a function of $\ln \xi$,
$\xi \equiv N \Delta$, with $N \le 30$ and $10^{-3}\le \Delta \le 1$.
The black curve is a best fit with functional form 
$\gamma(\xi) \propto \xi^{-b} \exp (- a \ln^2 \xi)$.
}
\end{figure}

The values of the optimal squeezing fraction $\beta_{\rm opt}=\beta_{\rm
opt}(N,\Delta)$ and of the corresponding QFI $H(N,\beta_{\rm
opt},\Delta)$ have been numerically evaluated and are reported in Fig.
\ref{f:optsqueezing} and Fig. \ref{f:QFIPhDiffCV} respectively.  As we
can see in Fig. \ref{f:optsqueezing}, for a low level 
noise the squeezing fraction is almost equal to one. In particular,
in each plot, for the lowest value of $\Delta$, we obtain $\beta_{\rm
opt}(N,\Delta)=1$ and thus the optimal probe state is the squeezed
vacuum state, as it happens in the noiseless case \cite{Mon06}.
As far as the noise $\Delta$ increases the squeezing
fraction decreases as a function of the average number of photons.
This means that for increasing values of the noise and of the energy, 
it is more convenient to employ the energy in increasing the coherent 
amplitude rather than the squeezing of the probe. 
Let us now focus on the behavior of the QFI $H(N,\beta_{\rm
opt},\Delta)$. In the left panel of Fig. \ref{f:QFIPhDiffCV}, we 
report the typical behavior of the QFI as a function of $N$ and
for different values of $\Delta$. The QFI
increases by increasing the average photon number $N$, and decreases 
with the noise parameter $\Delta$. For the lowest value of $\Delta$, 
we also observe that the noiseless limit $H(N,\beta=1,\Delta=0) = 
8(N^2 + N)$ \cite{Mon06} is approached, at least for $N$ not too 
large.
\par
\begin{figure}[h!]
\includegraphics[width=0.45\columnwidth]{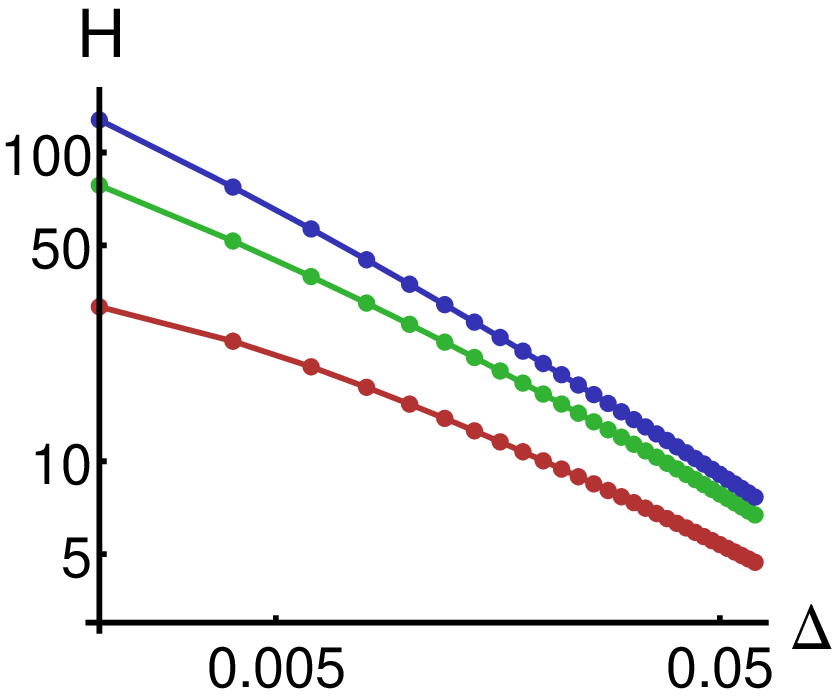}
\includegraphics[width=0.45\columnwidth]{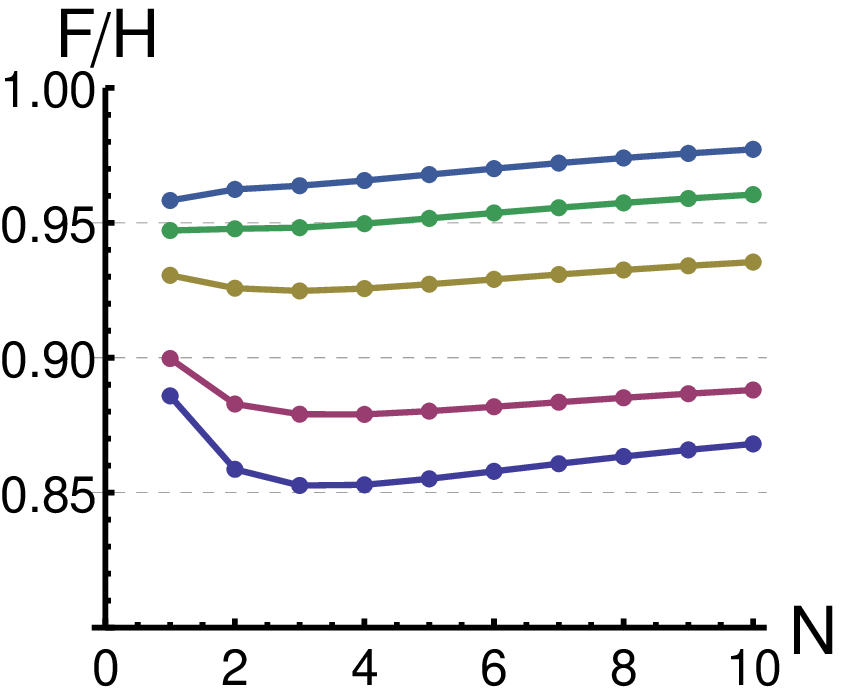}
\caption{
\label{f:f3} (Color online)
(Left): Log-log plot of the QFI for 
optimized pure input Gaussian states as a function of the noise 
parameter $\Delta$ for different values of the average photon number. 
From bottom to top: $N=\{2,5,10\}$. (Right):
Ratio between the Fisher Information of homodyne detection
on coherent states and the corresponding QFI, as a function of the
number of photons of the probe states and for different values of $\Delta$.
From bottom to top: $\Delta^2=\{ 0.5, 1.0, 1.5, 2.0, 5\}$.}
\end{figure}

As we have already mentioned above, because of the form of the
phase-diffusion map, we expect that the product $\xi = N\Delta$ 
plays a role in the estimation properties. In fact, by exploring 
a large range of values for $N$ and $\Delta$ a scaling law 
emerges from numerical analysis, which may be written as
\begin{equation}
H(N,\Delta) \simeq k^2 H(N/k, k\Delta)\,. \label{eq:LoSF}
\end{equation}
That is, $H(N,\Delta)=N/\Delta\, \gamma(\xi)=N^2
\gamma(\xi)/\xi=\xi\gamma (\xi)/\Delta^2$ where $0<\gamma(\xi)<1$ 
is a universal function independent on $\Delta$ and $N$. The larger is $\xi$
the more accurate is the scaling law. If we fix $\xi$ the QFI for
different pairs of $N$ and $\Delta$ have the same value, up to a
rescaling by a factor $k^2$, where $k$ is the ratio between the two
average photon numbers, or  equivalently of the two noise parameters.
The scaling is illustrated in the right panel of Fig.
\ref{f:QFIPhDiffCV} where we report the quantity $-\ln\gamma(\xi)$ as a
function of $\ln \xi$ (orange points) together with a two-parameter fit
(black curve) of the form $\gamma(\xi) \propto \xi^{-b} \exp (- a \ln^2
\xi)$, that provides a good representation of data. Using the above
results, the quantum Cram\`er-Rao bound for the precision of an
optimal estimator of the phase-shift may be written as 
$\hbox{Var}(\phi) \gtrsim \frac{\Delta}{\gamma(\xi)
N}=\frac{\xi}{\gamma(\xi)N^2}$. For small values of $\xi$ 
the quantity $\xi\gamma(\xi)$ is of  order of unity and thus
Heisenberg limit $\hbox{Var}(\phi) \sim N^{-2}$ in precision 
may be achieved \cite{Lane93}.
We also found that another scaling law, though less accurate, holds 
also for the optimal squeezing fraction
\begin{align}
\beta_{\rm opt} (N,\Delta) \simeq \beta_{\rm opt} (N/k,k\Delta).
\end{align}
Though based on a physical and mathematical justification, we cannot 
expect these  scaling to be exact due to the non-Gaussianity of the 
state. However they give a useful and practical receipt to compare 
and predict phase estimation performances in different regimes of energy 
and noise. 
In the left panel of Fig. \ref{f:f3} we show the behavior of the quantum 
Fisher information at fixed average photon number as a function of $\Delta$. 
We notice that the $H(N,\Delta)$ is decreasing exponentially with the phase 
noise and that higher values of $N$ correspond to higher values of $H$.
\par
\begin{figure}[h!]
\includegraphics[width=0.45\columnwidth]{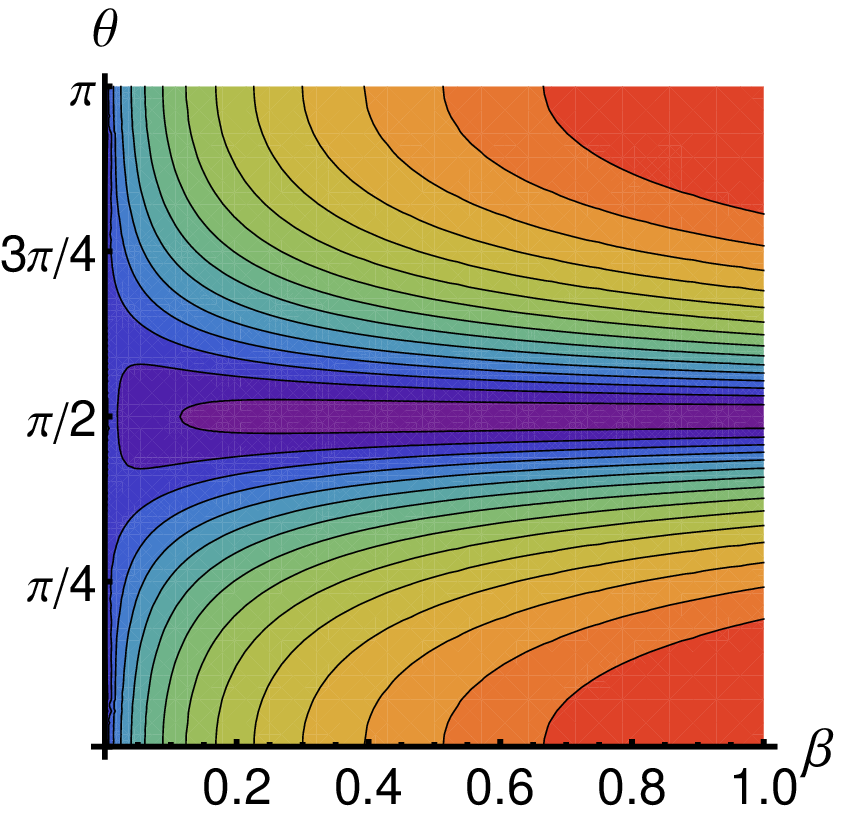}
\includegraphics[width=0.45\columnwidth]{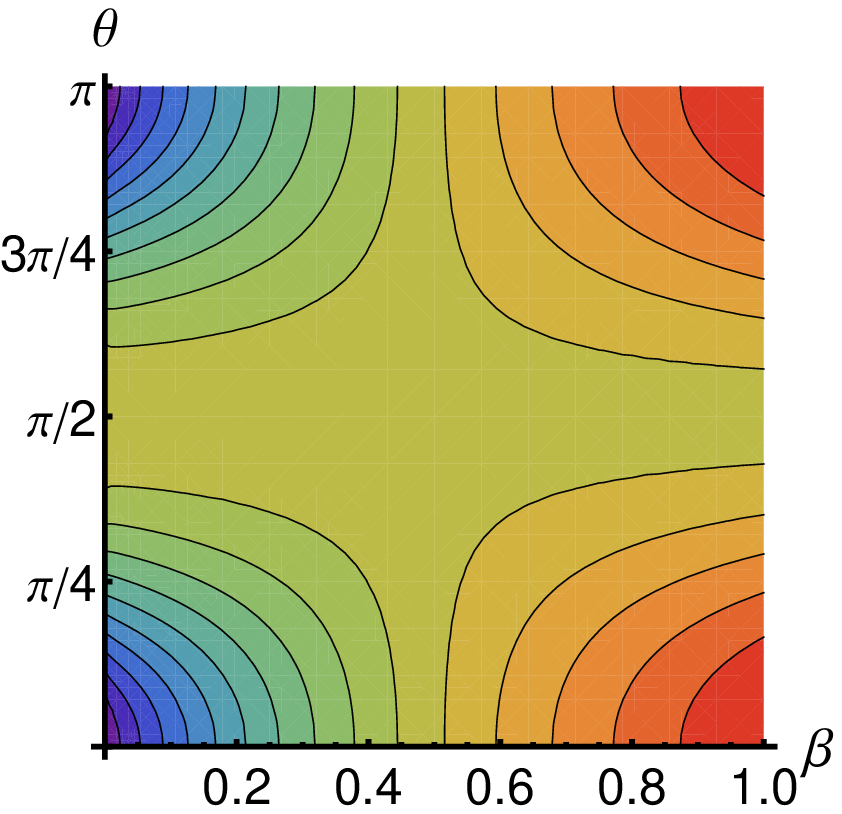}
\includegraphics[width=0.45\columnwidth]{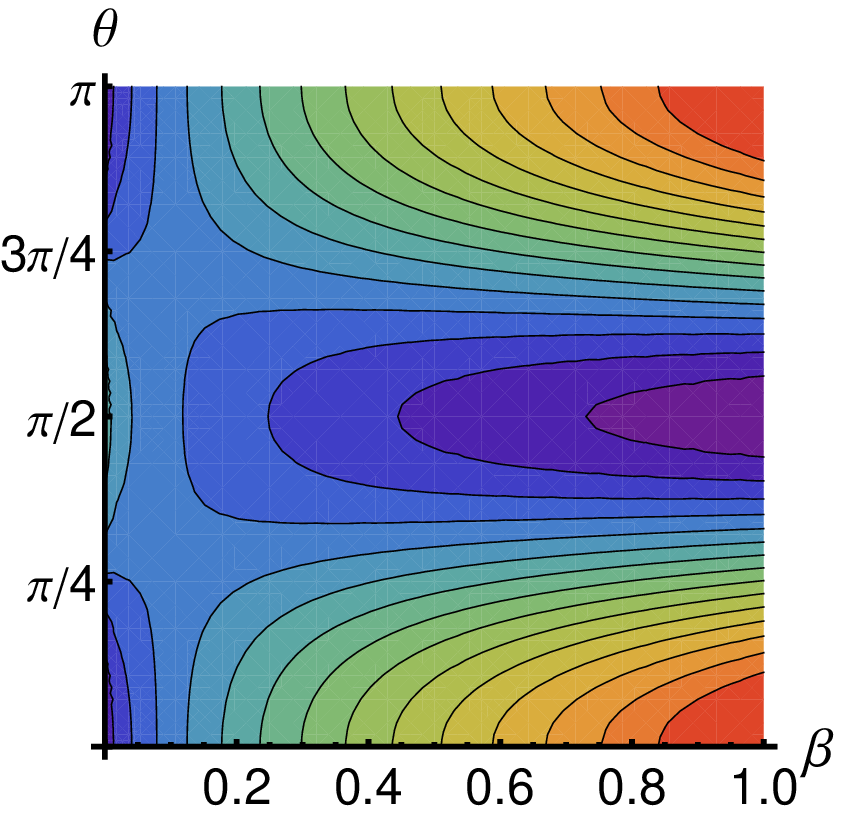}
\includegraphics[width=0.45\columnwidth]{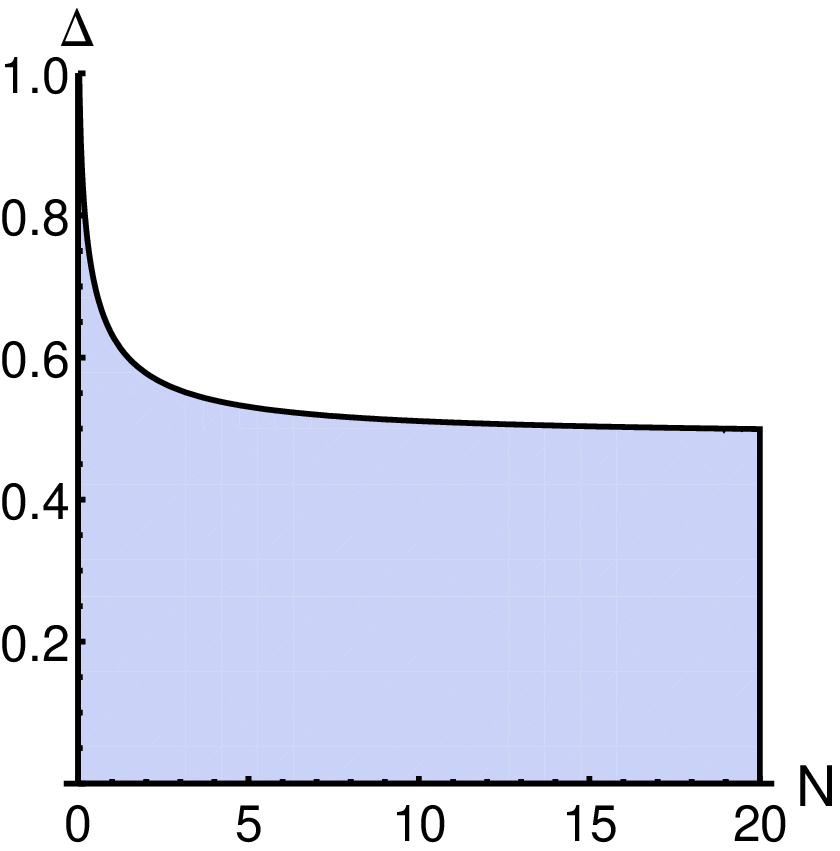}
\caption{
\label{f:PhEstQuadrature} (Color online)
Quadrature fluctuations $\Delta X_\theta^2$ as a function of 
the squeezing fraction of the input state $\beta$ and of 
the phase $\theta$ for different values of the noise amplitude 
$\Delta$ and the overall energy $N$.
Top left: $N=10$ and $\Delta=0.1$; top right: $N=10$ and $\Delta=0.6$; 
bottom left: $N=0.1$ and $\Delta=0.1$. Darker regions corresponds to 
smaller $\Delta X_\theta^2$. The plot in the bottom right panel 
illustrates the threshold $\Delta^*(N)$ between the two regions
where minimum fluctuations are achieved for $\beta=1$, $\theta=\pi/2$ 
(gray area) and $\beta=0$, $\theta=0$ respectively.} 
\end{figure}

In the noiseless case ($\Delta=0$) homodyne detection performed on input
squeezed vacuum states is optimal \cite{Mon06}, that is, its Fisher
information $F$ is equal to the QFI, $H(N)=8(N^2+N)$. A question thus
arises on whether this results also holds in the presence of phase
diffusion. Our numerical findings shows that this is true for a very
small amount of noise, i.e. $\Delta \ll 1$, whereas for increasing
$\Delta$ the ratio $F/H$ is moving away from unity quite quickly. On the
other hand, one can see that for high values of $\Delta$, basically when
coherent states are the optimal probe states maximizing the QFI,
homodyne detection of the quadrature $X=(a+a^{\dag})/2$ is again nearly
optimal, i.e. its Fisher information is again approaching the value of
the QFI evaluated in same conditions.  In the right panel of Fig. \ref{f:f3} we
plot the ratio between the Fisher information of homodyne detection and
the corresponding QFI: by increasing the noise $\Delta$ the ratio 
increases towards optimality ($F/H = 1$).  
This may understood looking at the behavior of quadrature fluctuations 
$\Delta X_{\theta}^2 = \langle X_\theta^2 \rangle - \langle X_\theta
\rangle ^2$ since the smaller is $\Delta X_{\theta}^2$ for a certain 
quadrature $X_\theta$, the more precise is the estimation of the
phase-shift through this quadrature.
In Fig.\ref{f:PhEstQuadrature}, we report a contour plot of 
$\log\Delta X_{\theta}^2$ as a function of the squeezing fraction 
of the input state $\beta$ and the quadrature phase $\theta$ for 
different values of $\Delta$ and of the overall energy $N$.  We 
see that for low noise, i.e. $\Delta\ll 1$, minimum fluctuations are 
obtained for the quadrature $\theta=\pi/2$ and for a squeezed vacuum 
state ($\beta=1$), whereas after a certain energy-dependent threshold 
level of noise $\Delta^*\equiv \Delta^*(N)$, we have a jump and the minimum 
fluctuations are achieved by measuring the $X$ quadrature ($\theta=0$) on 
coherent probes ($\beta=0$). 
This behavior is different compared to the behavior we have obtained 
for the QFI, see Fig.  \ref{f:optsqueezing}. There, for intermediate 
values of $\Delta$, the optimal squeezing fraction decreases monotonically 
from $\beta=1$ to $\beta=0$, whereas here we have only the extreme values. 
This exactly corresponds to the result discussed above: homodyne detection, as far as
we tune accordingly the measured quadrature, is optimal for very low
noise with squeezed vacuum probes ($\beta=0$), and for
large noise with coherent probes ($\beta=1$), while for
intermediate values of $\Delta$ homodyne detection is far from
optimality. Overall, we have that homodyne detection provides nearly
optimal phase estimation in the presence of either very small or large
phase diffusion, whereas it is still an open problem to find a feasible
measurement attaining the ultimate precision for a generic value of the
phase-diffusion noise parameter $\Delta$. 
 \par
In conclusion, we have attacked for the first time the problem of finding 
the optimal way to estimate a phase-shift in the presence of phase diffusion 
and we have obtained the ultimate quantum limits to precision for phase-shifted 
Gaussian states. By an extensive numerical analysis we have obtained an approximate 
scaling laws for both the quantum Fisher information and the optimal squeezing 
fraction in terms of the overall total energy and the amount of noise.
We also found that homodyne detection is a nearly optimal detection scheme 
for very small or large noise. Our results goes beyond the traditional
analysis of the quantum phase measurement problem and may be relevant
for the development of phase-shift keyed optical communication schemes
\cite{PSK06}.
\par
MGG acknowledges the UK EPSRC for financial support.

\end{document}